# Superoscillating response of a nonlinear system on a harmonic signal

D G Baranov,[1,2] A P Vinogradov,[1,2,3] and A A Lisyansky[4,5]

[1]Moscow Institute of Physics and Technology, 9 Institutskiy per., Dolgoprudny 141700, Russia
[2]All-Russia Research Institute of Automatics, 22 Sushchevskaya, Moscow 127055, Russia
[3]Institute for Theoretical and Applied Electromagnetics, 13 Izhorskaya, Moscow 125412, Russia
[4]Department of Physics, Queens College of the City University of New York, Queens, New York 11367, USA
[5]The Graduate Center of the City University of New York, New York, New York 10016, USA

**Abstract.** We demonstrate that a superoscillating in time signal may be obtained as a nonlinear response on a harmonic low-frequency input. Using the realization of a superoscillating function proposed by (Huang et al. 2007 *J. Opt. A: Pure Appl. Opt.* **9** S285-8) as an example, we synthesize the response function of such a nonlinear transformer and investigated its robustness with respect to the frequency and amplitude variations of the input signal.

Superoscillations is a counterintuitive mathematical effect, in which a band-limited function $f(t)$ (the function whose Fourier transform satisfies the condition $\hat{f}(\omega) = 0$ for all frequencies $|\omega| > \omega_{\max}$) may oscillate with a frequency much greater than $\omega_{\max}$. After the discovery of such functions by Berry [1], mathematical properties of superoscilalting functions have been studied in detail [2-7] and various mathematical approaches to their construction have been suggested [8, 9]. The concept of superoscillations has proven to be extremely fruitful in nanophotonics, where superoscillations enable deep subwavelength focusing of electromagnetic fields without use of evanescent waves [10-12]} (for review see Ref. [13]). More recently, superoscillations were studied in the time domain. Particularly, it has been shown that a quantum two-level emitter can be excited by a superoscillating electric field whose spectral components lie below the transition frequency of the emitter [14]. In another study, it has been found that a superoscillating electromagnetic signal can propagate trough absorbing media over length scales far exceeding the absorption length [15].

Here we explore a method of nonlinear synthesis of a superoscillating signal from a low-frequency single-harmonic input. We employ the technique of the harmonic synthesis [16] and explicitly construct the transformation function $f(z)$ which transforms a low-frequency harmonic $z(t)$ into a superoscillating function $y(t) = f(z(t))$.

Let us formulate the problem more rigorously. At the input of an inertialess nonlinear system we have a harmonic oscillation $z(t) = \cos \omega_0 t$. The output function $y(t)$ is expected to show a superoscillating behaviour. The problem is to find a nonlinear transformation function $f(z)$ which relates values of the input and output signals at the current time $t$. We are interested in the case when the output superoscillating function is represented as a superposition of $N$ harmonic oscillations which frequencies are multiple of $\omega_0$. An example of such a superoscillating function is presented in Ref. [10]:

$$y(t) = \sum_{n=0}^{5} A_n \cos n\omega_0 t, \quad A_0 = 1, \quad A_1 = 13295000, \quad A_2 = -30802818, \quad A_3 = 26581909, \quad A_4 = -10836909,$$

$A_5 = 1762818$, $\omega_0 = 1$. This function is plotted in figure 1, where the fastest harmonic $\cos(5t)$ is also shown for comparison. It is clearly seen that in time interval $-0.1 < t < 0.1$ the superoscillating function $y(t)$ is well approximated by function $f_{app}(t) = (\cos 43t + 1)/2$, which oscillates nearly 9 times faster than the cut-off component.

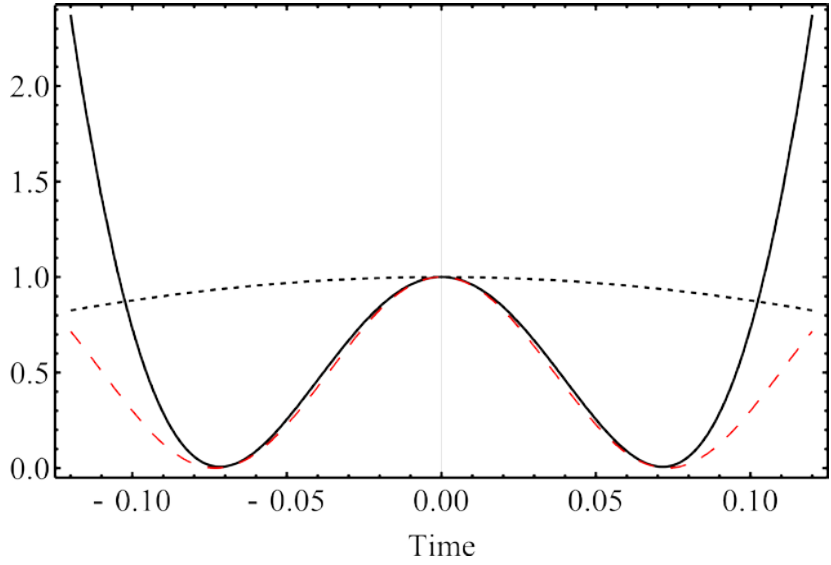

**Figure 1** Superoscillating function $y(t)$ containing spectral components $\omega_n = n$ (solid line) and the fastest component with frequency $\omega_5 = 5$ (black dashed line). The red dashed line shows the approximation of $y(t)$ near $t = 0$.

We seek for the desired transformation function in the form of a polynomial:

$$f(z) = a_0 + a_1 z + \ldots + a_N z^N. \tag{1}$$

Then, equality $y(t) = f(z(t))$ can be recast in the form

$$\sum_{n=0}^{N} a_n \cos^n \omega_0 t = \sum_{n=0}^{N} A_n \cos n\omega_0 t. \tag{2}$$

Having the set of values $A_n$ which yields the superoscillating feature of $y(t)$, we can establish general expression for $a_n$, which constitutes the transformation. Below we show explicit expressions for the case of $N = 6$ harmonic components of the output signal, which can be found in Ref. [16]. It is convenient to write expressions for the odd and even coefficients separately:

$$\begin{aligned}
a_0 &= 2^0 (A_0 - A_2 + A_4), \\
a_2 &= 2^1 (A_2 - 4A_4), \\
a_4 &= 2^3 (A_4),
\end{aligned} \tag{3a}$$

$$\begin{aligned}
a_1 &= 2^0 (A_1 - 3A_3 + 5A_5), \\
a_3 &= 2^2 (A_3 - 5A_5), \\
a_5 &= 2^4 (A_5).
\end{aligned} \tag{3b}$$

These formulas can be generalized for a greater number of components using rules developed in Refs. [16, 17].

Equations (3a) and (3b) together with equation (1) determine the desired transformation function $f(z)$ of the non-delay system. The resulting transformation for the given superoscillating function takes the form:

$$f(z) = 10^7 (1.9965910 - 5.7636637z - 2.5089636z^2 \\ + 7.1071276z^3 - 8.6695272z^4 - 2.8205088z^5)$$

(4)

Nonlinear characteristic of this transformation yielding superoscillating output is shown in figure 2. For the range of the input signal values $-1 \leq z \leq 1$ the output signal lies in the 7 orders of magnitude broader range. However, such extreme amplification may be eliminated by scaling down the transformation by a factor of $10^7$ [see equation (4)]. This decreases the superoscillations amplitude, but leaves the shape of the superoscillating function unchanged. One of features of the nonlinear transformation visible from figure 2 is that $f(0) \neq 0$, i.e., the current form of the transformation requires non-zero response of the system on zero input. Obviously, this is an unphysical property of the system. In order to fix this, one can subtract the constant $a_0$ from the nonlinear transformation $f(z)$. Obviously, the shifted output $y(t) - a_0$ is still superoscillating, but the resulting system with the transformation $F(z) = f(z) - a_0$ does not generate any signal at zero input.

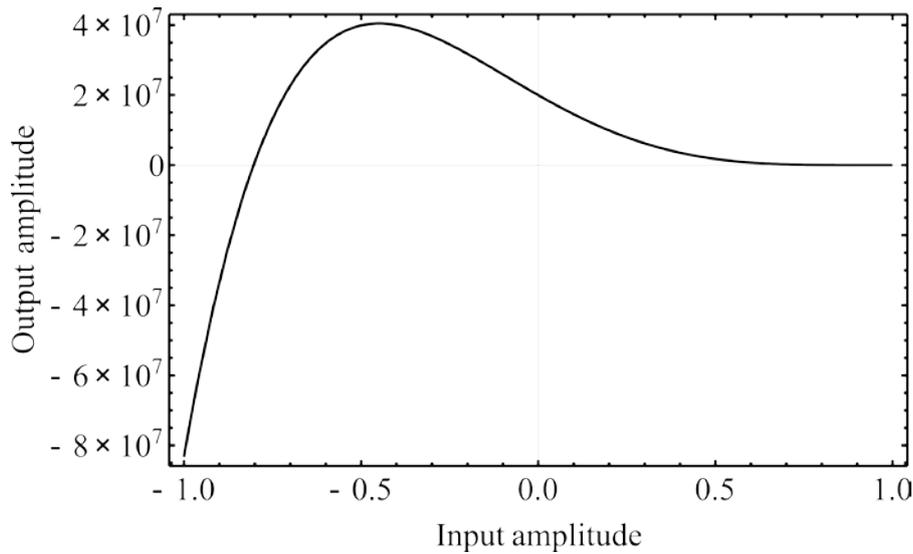

**Figure 2** Input-output characteristic of nonlinear transformation (4).

Now let us briefly discuss how robust the obtained transformation is against a variation of the input signal. Firstly, we note that transformation (4) is frequency scalable, i.e., if the input signal is given by $z_\alpha(t) = \cos \alpha \omega_0 t$, then one would obtain the output in the form $y_\alpha(t) = y(\alpha t)$, so that the superoscillating behaviour of the output signal is preserved.

Figure 3 shows the output generated by the transformation $f(z)$ for different amplitudes $z_0$ of the input signal $z(t) = z_0 \cos \omega_0 t$. Due to its nonlinear character, transformation (4) distorts the signal when its amplitude differs from the reference value $z_0 = 1$ for which the initial transformation is designed by equations (1) – (4). For 1% increase of the input amplitude, the output is still superoscillating while the distance between the neighboring minima increases (orange and red curves).The phenomenon is more sensitive, however, to a decrease of the input amplitude: its 1% variation drastically changes shape of the output function and kills superoscillations (the green curve). Overall, the current form of the nonlinear transformation is tolerant to ~ 0.5% variation of the input signal amplitude.

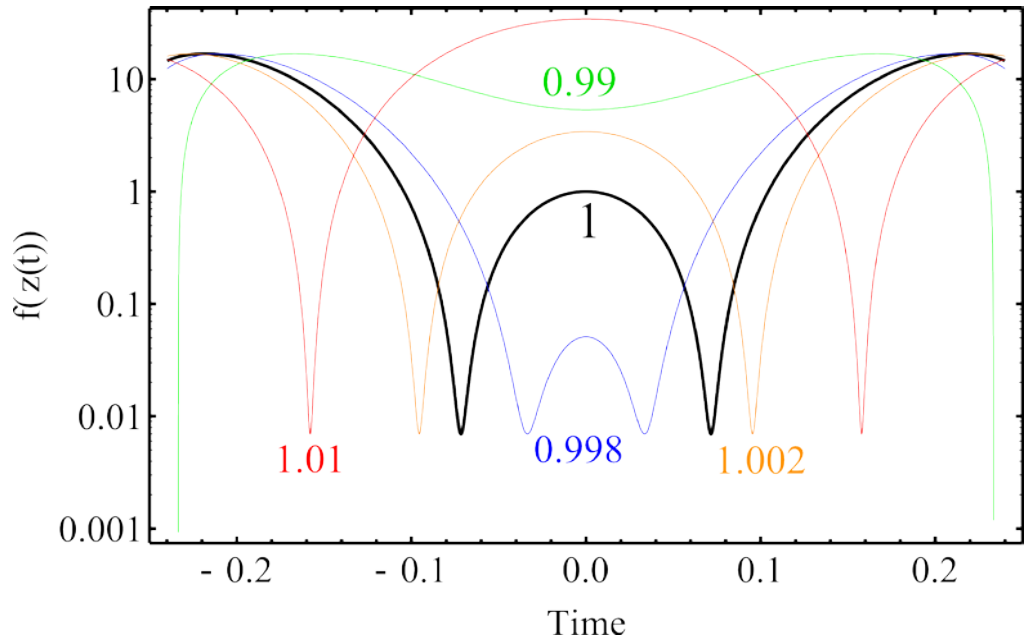

**Figure 3** Effect of variations of the input signal amplitude on the shape of the output signal $y(t)$. Black thick curve shows the output signal for the perfectly matched input amplitude $z_0 = 1$ also shown in figure 1. Numbers at the curves indicate the values of the input signal amplitude.

To conclude, we have demonstrated synthesis of a superoscillating function from a single-frequency input signal in a generic nonlinear inertialess system. We have derived an expression for the system transformation function which performs such synthesis and discussed robustness of superoscillation synthesis against variations of the input signal.

**Acknowledgements**

The work was supported by RFBR grants No. 13-02-92660, by Dynasty Foundation, by the NSF under Grant No. DMR-1312707, and by PSC-CUNY research award.